# Electric-field control of two-dimensional ferromagnetic properties by chiral ionic gating

Hideki Matsuoka,[1, *] Amaki Moriyama[1, 2], Tomohiro Hori[1, 3], Yoshinori Tokura[3, 4, 5], Yoshihiro Iwasa[3, 4], Shu Seki[2], Masayuki Suda[2, 6], Naoya Kanazawa[1]

1. Institute of Industrial Science, The University of Tokyo, Tokyo 153-8505, Japan
2. Department of Molecular Engineering, Graduate School of Engineering, Kyoto University, Kyoto 615-8510, Japan
3. Department of Applied Physics, School of Engineering, The University of Tokyo, Tokyo 113-8656, Japan
4. Center for Emergent Matter Science, RIKEN, Wako, Saitama 351-0198, Japan
5. Tokyo College, The University of Tokyo, Tokyo 113-8656, Japan
6. Department of Chemistry, Graduate School of Science, Nagoya University, Nagoya 464-8602, Japan

**ABSTRACT**.

Chiral molecular systems offer unique pathways to control spin and magnetism beyond conventional symmetry operations. Here, we demonstrate that chiral ionic liquids enable electric-field modulation of two-dimensional (2D) ferromagnetism in FeSi(111) thin films via electric double-layer transistor (EDLT) gating. FeSi hosts chemically-stable, surface-confined ferromagnetism without bulk moments, making the interfacial spins highly responsive to chiral-ion adsorption. Using both achiral and chiral ionic liquids, we systematically compare electrochemical and electrostatic gating effects. While both gating modes modulate magnetic properties such as anomalous Hall conductivity and coercive field, only chiral ionic gating biases the ratio of up- and down-magnetized domains in a handedness-dependent manner, evidencing chirality-induced symmetry breaking. This work establishes chiral ion gating as a novel strategy for controlling magnetic order and opens new directions for chiral spintronics.

**Text**

Electrical control of magnetic states is not only indispensable for realizing next-generation spintronic devices but also powerful for probing the fundamental physics of exotic magnets[1]. Traditionally, electric-field gating that modulates carrier density has long been employed to tune key magnetic parameters such as Curie temperature ($T_C$), coercive field ($H_c$), magnetic anisotropy, and magnetic transport properties, mainly in magnetic semiconductors[2–4]. More recently, this approach has been extended to magnetic oxides[5], magnetic topological insulators[6], and van der Waals magnets[7–11], providing a versatile pathway for investigating the interplay between magnetism and band-structure singularities and topological features.

These advances in electrical modulation of magnetic states are closely linked to the progress in gating techniques, as represented by ionic gating techniques[12,13]. Ionic gating is derived from conventional field-effect transistor (FET) architectures by replacing the solid dielectric layer with an electrolyte, often an ionic liquid (Figure 1(a)). This technique generally operates in two distinct modes. The first involves electrochemical reactions, characterized by atomic migration driven by redox processes (Figure 1(b)). For example, lithium-ion intercalation into $Fe_xGeTe_2$ enables room-temperature ferromagnetism[7,10], whereas proton or oxygen insertion/removal in cobalt oxides switches among three magnetic phases[14]. The second mode, termed electric double-layer transistor (EDLT), involves electrostatic carrier doping without electrochemical reactions (Figure 1(c)). Here, a substantial capacitance at the liquid-sample interfaces enables the doping of carriers with a scale one order of magnitude greater than traditional FETs, yielding pronounced changes in magnetic states in atomically thin films of various materials, including diluted magnetic semiconductors (Ga,Mn)As[15], magnetic oxides Co-doped $TiO_2$[5], magnetic topological insulators Mn-doped $Bi_2Te_{3-y}Se_y$[6], and a van der Waals magnet $Cr_2Ge_2Te_6$[9].

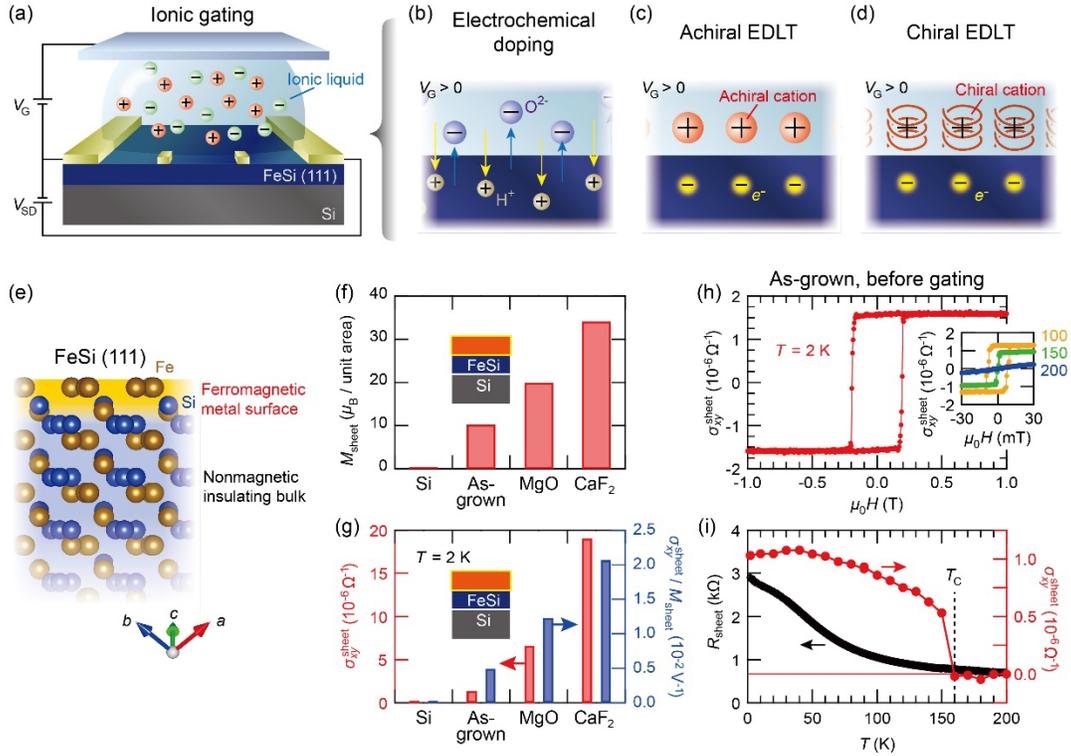

**Figure 1.**

**Basic properties and anomalous Hall effect of FeSi (111) thin films before ionic gating.**

(a) Schematic illustration of electric double-layer transistor (EDLT) using an ionic liquid. (b-d) Schematic comparison between three gating modes: (b) electrochemical doping, (c) EDLT with achiral ionic liquid, and (d) EDLT with chiral ionic liquid. (e) Crystal structure and magnetic state of FeSi (111) film with the ferromagnetic topological surface state. The crystal structure is drawn by VESTA[16]. (f-g) Comparison of (f) sheet magnetization $M^{\text{sheet}}$ and (g) sheet anomalous Hall conductivity $\sigma_{xy}^{\text{sheet}}$ between different capping layers including as-grown sample without capping layer. (h) Magnetic-field dependence of $\sigma_{xy}^{\text{sheet}}$ at various temperatures. (i) Temperature dependence of sheet resistance $R_{\text{sheet}}$ and spontaneous sheet anomalous Hall conductivity $\sigma_{xy}^{\text{sheet}}$ for a typical FeSi (111) thin film. The dashed line shows the Curie temperature $T_C$.

In this study, we introduce a novel mode and functionality within the ion-gating framework: chiral EDLT. In this approach, ionic liquids comprising chiral organic molecules form an electric double-layer interface at the sample surface (Figure 1(d)). While conventional EDLT has typically utilized achiral ionic liquids (*e.g.*, [DEME][TFSI]), the development of chiral ionic liquids incorporating chiral organic cations has recently opened additional possibilities[17,18]. The integration of chirality into spintronic systems has gained particular attention in the context of the chirality-induced spin selectivity (CISS) effect, where the quantum coupling between electronic spin and molecular chirality enables diverse spin-related functionalities ranging from efficient charge-spin conversion to enantioselectivity[19–26]. Inspired by the CISS effect, the incorporation of chiral organic molecules into ion-gated magnetic systems holds the potential to realize such multifaceted functionalities beyond conventional tuning of $T_\mathrm{C}$ or magnetic anisotropy. By using the intrinsic chirality of the gating medium, we aim to uncover symmetry-breaking effects and nontrivial magnetic responses, paving the way for chiral spintronic devices.

Our study utilizes FeSi (111) epitaxial thin films as the magnetic material subjected to gating. While FeSi is known to be the narrow-gap insulator in bulk form[27–31], its surface exhibits metallic conduction[32–34], together with quasi-two-dimensional (2D) ferromagnetism and strong spin-orbit-coupled bands that are all rooted in a nontrivial Zak phase and emerge only at the surface[32,35–37] (Figure 1(e)). Consequently, the surface-derived nature of this magnetism manifests in the strong sensitivity of properties such as sheet magnetization $M_\mathrm{sheet}$, anomalous sheet Hall conductivity $\sigma_{xy}^\mathrm{sheet}$, and anomalous Hall coefficient $\sigma_{xy}^\mathrm{sheet} / M_\mathrm{sheet}$ to the choice of capping material (Figure 1(f) and 1(g))[36]. Notably, the dependence on capping materials follows a systematic trend: interfaces that better preserve the Zak-phase-driven surface polarization through the reduced interfacial hybridization yield larger $M_\mathrm{sheet}$, $\sigma_{xy}^\mathrm{sheet}$, and $\sigma_{xy}^\mathrm{sheet}/M_\mathrm{sheet}$. This surface sensitivity suggests that surface-specific modulation techniques, such as EDLTs, are particularly effective for tuning the magnetism in FeSi. In this research, we systematically apply three ion-gating modes, *i.e.*, electrochemical reactions, achiral ionic gating and chiral ionic gating, to the FeSi (111) thin films, aiming to modulate their magnetic properties dramatically.

Initially, we grew FeSi (111) thin films without any capping layer (hereafter *as-grown* FeSi) and patterned them into gate devices to verify their ferromagnetic state before applying ionic gating (Supplemental Materials). Figure 1(h) illustrates the magnetic-field (*H*) dependence of $\sigma_{xy}^{\text{sheet}}$ at various temperatures, and Figure 1(i) presents the temperature (*T*) dependence of sheet resistance $R_{\text{sheet}}$. Temperature dependence of $R_{\text{sheet}}$ shows an insulating behavior, while it remains measurably conductive down to *T* = 2 K. Clear anomalous Hall effect (AHE) signals appear below $T_{\text{C}}$ with *H*-hysteresis, indicative of perpendicular magnetic anisotropy. The AHE signal disappears above 200 K, and a detailed zero-field AHE measurement gives a Curie temperature $T_{\text{C}}$ = 160 K. As shown in Figures 1(f) and 1(g), the magnetism of the as-grown sample is partially suppressed due to the presence of a naturally formed surface oxide layer with a thickness of approximately 1 nm[32]. Nevertheless, the EDLT approach is expected to modulate the surface magnetism even through this ultrathin insulating layer.

To investigate the influence of ionic gating, we employed the achiral ionic liquid DEME-TFSI (Figure 2(a)) and utilized two different temperature windows to separate electrostatic EDLT operation from electrochemical doping (also see Figure S1 for comparing the results in different modes and conditions). At higher temperatures, chemical reactions are thermally activated, promoting electrochemical processes upon gating as well. According to prior studies, electrochemical gating is performed at *T* > 300 K[38–40], whereas electrostatic carrier accumulation (EDLT mode) is typically carried out at temperatures slightly above the freezing point of the ionic liquid (~ 200 K for DEME-TFSI)[41–44], around $T \sim 220$ K.

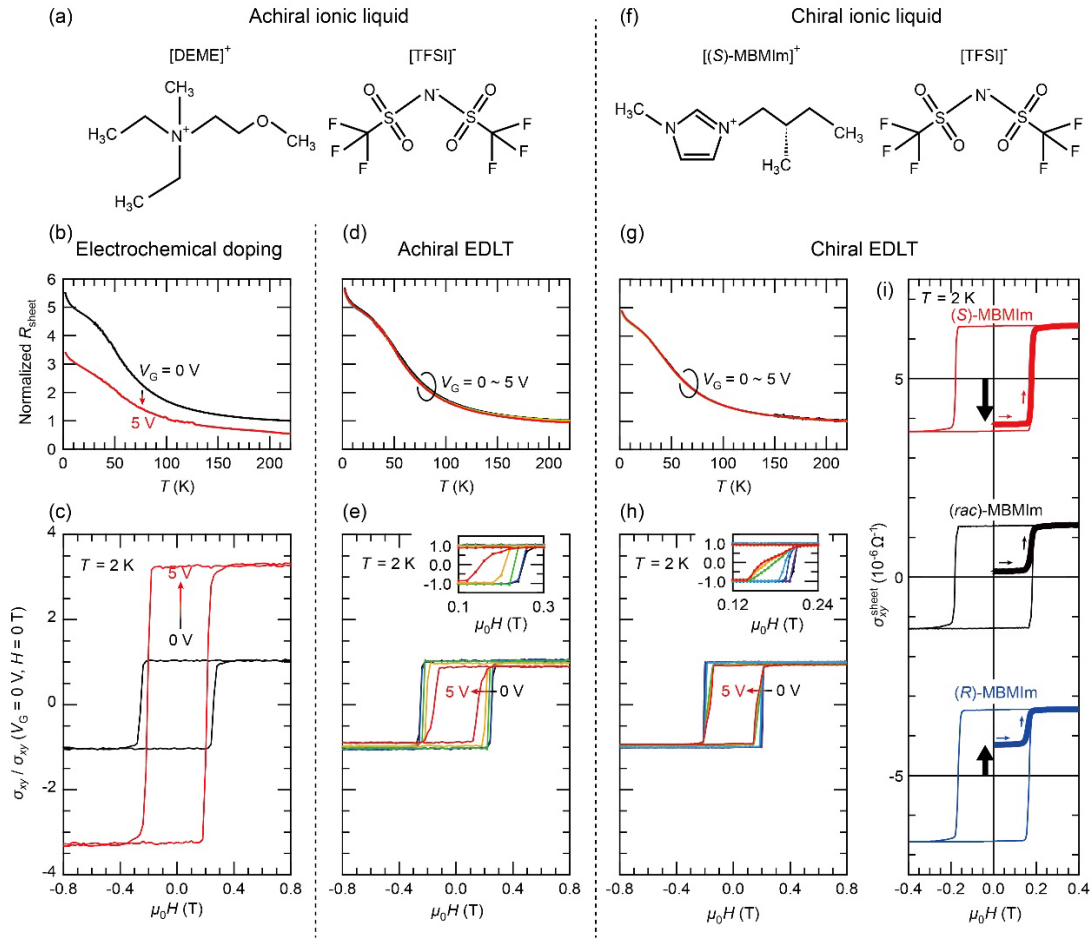

**Figure 2.**

**Ionic gating effects on the transport properties of FeSi under different modes.**

(a) Molecular structure of the achiral ionic liquid [DEME][TFSI]. (b, c) Electrochemical doping: (b) temperature dependence of sheet resistance $R_{sheet}$ and (c) magnetic-field dependence of anomalous Hall conductivity $\sigma_{xy}$ at 2 K before and after applying gate voltage. (d, e) EDLT with achiral ionic liquid: (d) temperature dependence of $R_{sheet}$ and (e) magnetic-field dependence of $\sigma_{xy}$ at 2 K before and after applying gate voltage. (f) Molecular structure of the chiral ionic liquid [(S)-MBMIm][TFSI]. (g, h) EDLT with chiral ionic liquid: (g) temperature dependence of $R_{sheet}$ and (h) magnetic-field dependence of $\sigma_{xy}$ at 2 K before and after applying gate voltage. (i) Magnetic-field dependence of $\sigma_{xy}$ at 2 K for devices coated with ionic liquids of different chirality: (S)-MBMIm, racemic (rac)-MBMIm, and (R)-MBMIm. Thick curves indicate the initial sweep starting from zero-field. The initial $\sigma_{xy}$-H sweep under zero gate voltage ($V_G = 0$ V) after ZFC reveals distinct domain imbalance only under the chiral ionic liquid, indicating chirality-induced magnetic domain polarization.

We first examine the results obtained in the electrochemical-doping mode. Figures 2(b) and 2(c) show the $T$-dependence of $R_{sheet}$ and the $H$-dependence of $\sigma_{xy}^{sheet}$ after high-temperature gating under a gate voltage $V_G = +5$V. Although the resistance change is significant, it involves both contributions from bulk and surface conduction channels, making it difficult to pinpoint a specific origin. Because the magnitude of this change exceeds the variation reported in the previous study on capping-layer dependence[36], substantial carrier doping likely occurs in the bulk region.

In parallel, changes in the surface state are also recognized from variations in $\sigma_{xy}^{sheet}$, since the ferromagnetic order is confined to the FeSi surface. Electrochemical gating causes a notable decrease in $H_c$ and boosts $\sigma_{xy}^{sheet}$ by 3.1 times. This enhancement of AHE resembles the case of FeSi capped with MgO (Figure 1(g)), suggesting that electrochemical reactions altered the native surface oxidization. When the gate voltage $V_G$ is returned to zero, the resistance fully recovers and $\sigma_{xy}^{sheet}$ remains partially elevated (Figure S2). These contrasting behaviors indicate that both reversible and irreversible processes are involved in the electrochemical modulation mechanism. We will discuss the details of the mechanism in the discussion section later.

Next, we focus on the results obtained under the achiral EDLT mode. Figures 2(d) and 2(e) present the $T$-dependence of $R_{sheet}$ and the $H$-dependence of $\sigma_{xy}^{sheet}$, respectively. Across the entire temperature range, the resistance remains essentially unchanged, indicating that electrostatic carrier accumulation has little impact on the already highly metallic FeSi surface (see Figure S3 for carrier density variation versus $V_G$). In contrast, significant modulation is observed in the AHE: the coercive field $H_c$ decreases by 40 % and the amplitude of $\sigma_{xy}^{sheet}$ drops by 14 % with applying $V_G$ = +5 V. After removing the $V_G$ to check reversibility, the $H_c$ fully returns to its original state (see Figure S1), consistent with the static and reversible nature of the EDLT mechanism.

We now present the results obtained under the chiral EDLT mode. Molecular structures of the ionic liquids used in the chiral EDLT modes are displayed in Figure 2(f), where the chiral ionic liquid contains the chiral cation [($S$ or $R$)-MBMIm]$^+$ paired with the same anion [TFSI]$^-$ as achiral ionic liquid [DEME][TFSI]. Figures 2(g) and 2(h) show the $T$-dependence of $R_{sheet}$ and the $H$-dependence of $\sigma_{xy}^{sheet}$, respectively. As in the achiral EDLT case, $R_{sheet}$ remains nearly unchanged across the entire temperature range, while the AHE shows a comparable response, with $H_c$ decreasing by 21% and $\sigma_{xy}^{sheet}$ by 10%. These results are qualitatively similar to those obtained with achiral EDLT, suggesting that the observed magnetic modulation originates from electric field effects or carrier doping rather than molecular chirality.

A key finding is the emergence of magnetic domain polarization under the chiral EDLT even without applying a gate voltage. We quantified this effect by measuring $\sigma_{xy}^{sheet}$ at $T$ = 2 K after zero-field cooling (ZFC), a protocol that reveals any imbalance between up- and down-moment domains under zero external magnetic field. Figure 2(i) displays the $\sigma_{xy}^{\text{sheet}}$-$H$ curves, highlighting the initial $H$-sweep from zero to positive values, under different conditions. Normally, ZFC leaves the up- and down-moment domains energetically equivalent, yielding nearly equal populations and therefore no net AHE signal. This is confirmed in the as-grown (*i.e.*, ionic liquid-free) device, where the domain polarization is absent (highlighted panel in Figure S4(b)). Here, the tiny offset from zero is attributed to residual stray magnetic fields in the measurement setup (see Supplemental Materials for the details of zero-field adjustment procedure). In contrast, when a chiral ionic liquid contacts with the sample surface under zero gate voltage ($V_G = 0$ V), a clear domain-polarization offset appears whose sign reverses with molecular handedness: down-moment domains for the *S*-enantiomer (top panel of Figure 2(i)); up-moment domains for the *R*-enantiomer (bottom panel of Figure 2(i)). Such a distinct offset disappears when a racemic mixture (namely, equal numbers of *S*- and *R*-type cations) is used, indicating that chirality of the liquid is responsible for the phenomenon (middle panel of Figure 2(i)).

Based on the experimental results presented above, we now discuss the underlying mechanisms behind the observed modulation of magnetic state under different gating modes. Figures 3(a–c) summarize the $V_G$-dependence of key physical quantities, $R_{\text{sheet}}$ ($T = 220$ K), $H_c$ ($T = 2$ K), and $\sigma_{xy}^{\text{sheet}}$ ($T = 2$ K), during the electrochemical doping, achiral EDLT, and chiral EDLT modes. In addition, a qualitative overview of the gating-induced changes and the domain polarization is provided in Table 1.

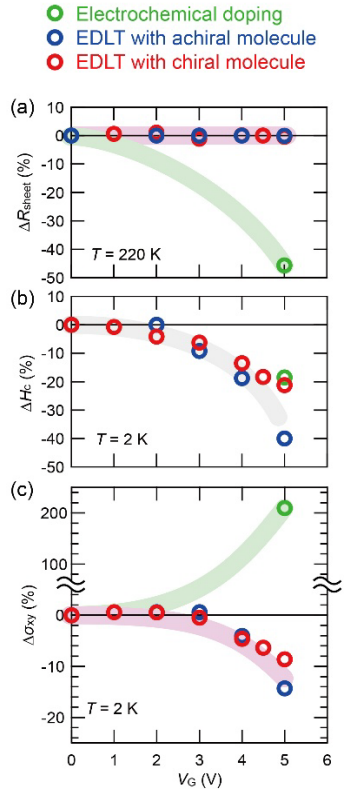

**Figure 3.**

**Gate-voltage-dependent modulation of key magnetic and transport parameters under different ionic gating modes.**

(a) Change in sheet resistance $\Delta R_{sheet}$ at 220 K. (b) Change in coercive field $\Delta H_c$ at 2 K. (c) Change in anomalous Hall conductivity $\Delta\sigma_{xy}$ at 2 K. Each parameter is shown for three gating modes: electrochemical doping (red), electric double-layer transistor (EDLT) with achiral ionic liquid (blue), and EDLT with chiral ionic liquid (green). All values are plotted as the percentage change relative to the initial (zero-bias) state.

| Parameter | Electrochemical doping ($V_G > 0$) | EDLT with achiral molecule ($V_G > 0$) | EDLT with chiral molecule ($V_G > 0$) |
|---|---|---|---|
| Sheet resistance $R_{sheet}$ | $\Delta < 0$ | $\Delta = 0$ | $\Delta = 0$ |
| Coercive field $H_c$ | $\Delta < 0$ | $\Delta < 0$ | $\Delta < 0$ |
| Anomalous Hall conductivity $\sigma_{xy}$ | $\Delta > 0$ | $\Delta < 0$ | $\Delta < 0$ |
| Magnetic domain control | no | no | yes |

**Table 1.**

**Summary of magnetic and transport responses of FeSi thin films under different ionic gating modes.**

The table lists the changes $\Delta$ in sheet resistance $R_{sheet}$, coercive field $H_c$, and anomalous Hall conductivity $\sigma_{xy}$ upon gate voltage application ($V_G > 0$) for three different gating modes: electrochemical doping, EDLT with achiral ionic liquid, and EDLT with chiral ionic liquid. Additionally, the presence or absence of zero-field magnetic domain polarization is indicated.

We first examine the origin of the similar modulations of magnetic state (decreases in $H_c$ and $\sigma_{xy}^{sheet}$) observed in both achiral and chiral EDLT modes. As shown in Figure 3, these two modes exhibit qualitatively identical trends, suggesting that a common mechanism is linked primarily to carrier accumulation and/or strong interfacial electric field. In conventional gate-controlled magnetism, such modulations are typically attributed to variations in carrier density, consistent with the Stoner criterion and RKKY interactions, where magnetic properties scale with carrier concentration[4,7,9]. In the surface state of FeSi, hole carriers are dominant ($n_h = 4.0 \pm 0.2 \times 10^{15}$/cm$^2$ for $V_G = 0$ V), and the application of a positive $V_G$ is expected to reduce the carrier density (Figure S3). According to the Stoner criterion, this reduction in carrier density would lead to a suppression of magnetic ordering, possibly accounting for the observed decreases in both $H_c$ and $\sigma_{xy}^{\text{sheet}}$. On the other hand, a simple estimate (namely, converting the $V_G$-induced change in carrier density $\Delta n_h$ (Figure S3) into the expected change in magnetic moment $\Delta M$ under the assumption of linear proportionality $\Delta M \propto \Delta n_h$) shows that carrier accumulation alone cannot account for such large reduction in $\sigma_{xy}^{\text{sheet}}$ ($\propto M$). Still, more subtle carrier-mediated mechanisms cannot be ruled out.

An alternative explanation is that the electric field itself, rather than carrier density modulation, drives the modulation of magnetic state. As described in Figures 1(f) and 1(g), the surface ferromagnetism in FeSi is coupled to surface polarization, and a positive gate voltage acts to suppress this polarization. Consequently, a reduction in surface polarization could lead to the simultaneous decreases in both $H_c$ and $\sigma_{xy}^{\text{sheet}}$.

We next consider the electrochemical doping mode, which yields qualitatively distinct behaviors in comparison to EDLT-based gating. As shown in Figures 3(a) and 3(c), this mode causes a 46 % drop in $R_{\text{sheet}}$ and a 210 % rise in $\sigma_{xy}^{\text{sheet}}$. These pronounced changes suggest mechanisms fundamentally different from those in electrostatic gating. Two primary reduction reactions are plausible under this mode: (1) oxygen removal from the native surface oxide, and (2) proton intercalation into the bulk lattice. Ferromagnetism at FeSi (111) surface is highly sensitive to surface protection against oxidization, and the observed enhancement in $\sigma_{xy}^{\text{sheet}}$ closely resembles the case of MgO capping (Figure 1(g)), which supports the role of oxide removal in strengthening the surface ferromagnetism. In contrast, the sharp fall in $R_{sheet}$ persists up to 200 K, where the bulk conduction dominates, implying that proton intercalation lowers the bulk resistivity. Taken together, both reactions are likely to occur simultaneously, with oxygen removal being primarily responsible for enhanced ferromagnetism.

A particularly intriguing phenomenon observed in this study is the sizable magnetic domain polarization induced by chiral EDLT. While similar domain polarization effects have been previously demonstrated in Co thin films interfaced with chiral molecules yielding domain polarizations up to ~30%[45], our FeSi(111) devices exhibit a significantly higher degree of domain polarization. This striking enhancement may stem from the uniquely 2D ferromagnetism of FeSi, which is confined to its topological surface states and thus highly sensitive to interfacial interactions.

However, the magnitude of the domain polarization varies between samples and declined gradually over repeated measurement cycles. Figure 4(a) shows representative data from the first few cycles for a device coated with the *S*-enantiomer. The plots trace the AHE signal recorded after each thermal treatment, where the sample was heated above the glass transition temperature (200 K) of the ionic liquid and then cooled down. (Also see Figures S5 and S6 for the detailed measurement procedures.) Quantitatively, the domain polarization ratio $M/M_{sat}$, calculated by normalizing the zero-field AHE to its saturated value, reached as high as 87% at its maximum. Notably, this large imbalance does not arise in the very first cycle; it usually becomes most pronounced around the second or third cycle and then decays in subsequent cycles. During these measurements, both positive and negative gate voltages were applied at specific cycles; however, the domain polarization remained unaffected by the gate voltage (see Figure S5). Figure 4(b) shows the evolution of domain polarization as a function of cycle number for devices gated with ionic liquids of different chirality (*S*-, *R*-, and racemic forms). The device with the *S*-enantiomer shows the largest polarization ratio of down-moment domains overall, while the polarization ratio of up-moment domains reaches a smaller peak in roughly the same cycle range and exhibits greater scatter for the *R*-enantiomer. The device with the racemic mixture remains essentially zero for all cycles. Figure 4(c) presents a histogram that again confirms finite and opposite polarizations for the two chiral ionic liquids (*S*- and *R*-enantiomers) and no significant effect for the racemate. Nevertheless, the domain polarization exhibits considerable variability, depending strongly on the sample and measurement cycle (also see Figure S6). These observations imply that the coupling strength is highly sensitive to interface preparation, including oxide thickness, surface cleanliness, and thermal history. Refining these parameters may stabilize and further amplify this chirality-driven effect.

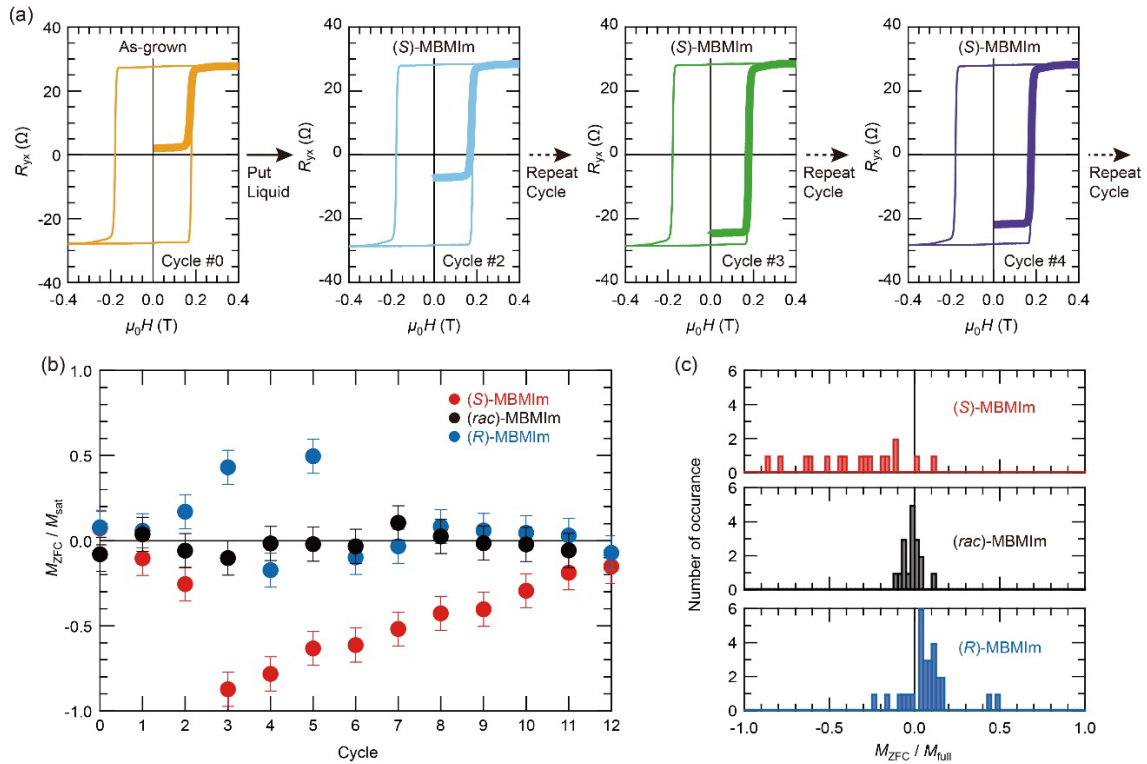

**Figure 4.**

**Magnetic domain polarization under chiral EDLT.**

(a) Evolution of the anomalous Hall resistance at $T = 2$ K over successive zero-field cooling (ZFC) cycles for a device gated with the *S*-enantiomer. Each panel corresponds to one cycle after the sample was heated above the ionic-liquid glass transition temperature and then zero-field cooled. The panels display only a few representative cycles, all measured at $V_G = 0$ V; additional cycles acquired at finite $V_G$ show no discernible gate-voltage effect on the domain polarization (Figure S5). (b) Domain polarization ratio $M_{ZFC}/M_{sat}$, estimated from zero-field $R_{yx}$, plotted as a function of cycle number for devices gated with the *S*-enantiomer (red), the *R*-enantiomer (blue), and the racemic mixture (black circles), respectively. (c) Histogram of domain polarization values collected from all devices and cycles for the *S*-enantiomer (top, red), the racemic mixture (middle, black), and the *R*-enantiomer (bottom, blue).

Only from a spatial symmetry perspective, the coexistence of molecular chirality and the surface-normal symmetry breaking inherent to the film-molecule interface may allow for an imbalance between up- and down-moment domains. Yet, lifting the degeneracy of these domains also requires time-reversal symmetry breaking. In conventional CISS, this is typically achieved via current flow. In other cases, such as enantioselective molecular reactions, adsorption/desorption kinetics may break time-reversal symmetry. In the present case, however, the ionic liquid is frozen below $T_C$ and both ionic motion and charge transport are strongly suppressed. This suggests that time-reversal symmetry breaking may arise from other mechanisms. One possibility is the recently proposed angular momentum flow associated with chiral phonons, which may induce spin polarization[46]. Another potential mechanism involves interfacial coupling between local vibrational modes of chiral molecules and surface spins. Notably, the observed domain polarization showed negligible dependence on gate voltage, suggesting that the effect is not dominated by local interfacial interactions but rather mediated by longer-range excitations such as phonons. Further investigation, such as phonon spectroscopy or spatially resolved imaging of domain structures, will be essential to unravel the microscopic origin of this unusually robust and efficient domain polarization.

In conclusion, we have demonstrated distinct modulation of the surface ferromagnetism in FeSi(111) thin films by employing ion gating with achiral and chiral ionic liquids. While electrochemical doping modulates both bulk and surface electronic states via redox-driven processes, electric double-layer gating with achiral and chiral ionic liquids induces notable changes in magnetic anisotropy and anomalous Hall conductivity, despite negligible changes in charge carrier density, suggesting a strong role of interfacial electric fields and spin-orbit coupling. Most remarkably, chiral ionic gating leads to spontaneous magnetic domain polarization under zero magnetic field, a symmetry-breaking effect not observed with racemic or achiral ionic liquids. This finding suggests a new functionality emergent from the interplay between molecular chirality and low-dimensional magnetism, potentially mediated by chiral phonons or interfacial dipole-spin coupling. These results expand the functionality of ionic gating toward chiral spin control and open new possibilities for designing symmetry-driven spintronic devices using molecularly engineered interfaces.

**Corresponding Author**

*Corresponding author. Email: hideki-m@iis.u-tokyo.ac.jp

**Author contributions**

T.H. and N.K. grew and characterized the FeSi thin films. H.M. and A.M. performed the gating experiments and analyzed the data. M.S. synthesized the chiral ionic liquids. S.S., M.S., and N.K. supervised the project. H.M. and N.K. wrote the manuscript. All authors discussed the results and commented on the manuscript.

**Notes**

The authors declare no competing financial interest.

**Acknowledgement**

This work was supported by JSPS KAKENHI (Grants No. 21K13888, No. 23H04017, No. 23H05431, No. 23H05462, No. 24H00417, No. 24H01212, No. 24H01652, No. 25H02126, 25H02141, and 25K22219), JST FOREST (Grants No. JPMJFR2038 and JPMJFR221V), JST CREST (Grant No. JPMJCR23O3), Mitsubishi Foundation, Sumitomo Foundation, and Tanaka Kikinzoku Memorial Foundation.